\documentclass{article}
\usepackage{amssymb}

\begin{document}

\title{Lyapunov Exponent for Pure and Random Fibonacci Chains}
\author{M. T. Velhinho and I. R. Pimentel$\dagger $\\
{\small Department of Physics and CFMC, University of Lisbon, Lisboa 1649, Portugal}}
\date{}
\maketitle

\begin{abstract}

We study the Lyapunov exponent for electron and phonon excitations, in pure
and random Fibonacci quasicrystal chains, using an exact real space
renormalization group method, which allows the calculation of the Lyapunov
exponent as a function of the energy. It is shown that the Lyapunov exponent
on a pure Fibonacci chain has a self-similar structure, characterized by a
scaling index that is independent of the energy for the electron
excitations, ''diagonal'' or ''off-diagonal'' quasiperiodic, but is a
function of the energy for the phonon excitations. This scaling behavior
implies the vanishing of the Lyapunov exponent for the states on the
spectrum, and hence the absence of localization on the Fibonacci chain, for
the various excitations considered. It is also shown that disordered
Fibonacci chains, with random tiling that introduces phason flips at certain
sites on the chain, exhibit the same Lyapunov exponent as the pure Fibonacci
chain, and hence this type of disorder is irrelevant, either in the case of
electron or phonon excitations.

\end{abstract}

\section{Introduction}

The experimental discovery of quasicrystals,$^{1}$ and also the building of
artificial multilayer structures by molecular beam epitaxy,$^{2}$ have
considerably stimulated the theoretical study of quasiperiodic systems.
$^{3-5}$ Quasicrystals have a deterministic aperiodicity that characterizes
them as intermediate structures between periodic crystals and disordered
materials, therefore being expected to display new behavior. There has been
in particular, great discussion on the nature of the energy spectrum and
eigenstates of electron and phonon excitations on quasicrystals. It is
questioned whether the spectrum is absolutely continuous, pointlike or
singular continuous, or correspondingly, if the states are extended,
localized or critical.

The Fibonacci chain is the simplest quasicrystal, a one-dimensional system
where the site or bond variables take one of the two values $A$ and $B$, and
are arranged in a Fibonacci sequence. The Fibonacci chain can be constructed
recursively by successive applications of a substitution rule, $A\rightarrow
AB$ and $B\rightarrow A$, or alternatively, by successive applications of a
concatenation rule, $S_{i}=S_{i-1}\otimes S_{i-2}$, $S_{i}$ being the
Fibonacci sequence at iteration $i$. The quasiperiodicity of the Fibonacci
chain is characterized by the golden mean $\tau =\left( 1+\sqrt{5}\right) /2$
, which gives the ratio of the number of $A$ and $B$ units. Tight-binding
electron and phonon excitations have been studied on a Fibonacci chain,
using mainly transfer-matrix$^{6-12}$ and real space renormalization group
techniques.$^{13-15}$ It has been found that the energy spectrum for these
excitations is a Cantor set with zero Lebesgue measure, this result having
in addition been proven$^{16}$ for the case of electronic excitations on a
site Fibonacci chain. The spectra of the periodic approximants to the
Fibonacci chain exhibit self-similarity in the band structure, with a
scaling index that for the electronic excitations is independent of the
energy, while for the phonon excitations it is a function of the energy.
$^{9} $ The integrated density of states for the various excitations presents
rich scaling behavior, with indices varying from the edge to the center of
the bands.$^{10-12,14}$The characterization of the eigenstates on a
Fibonacci chain is a more difficult task, and it has usually been restricted
to a few special energies on the spectrum, for which the states are found to
be self-similar or chaotic. More generally, it has been found evidence for
the states being neither extended nor localized in the usual sense.
$^{10-12}$

The localization properties of the states can be studied through the
calculation of the Lyapunov exponent $\gamma $, which characterizes the
evolution of a wavefunction along the chain.$^{17-19}$ The Lyapunov exponent
is zero for an extended or critical state, but is positive for a localized
state representing then the inverse of the localization length. Delyon and
Petritis$^{20}$ have proved that the Lyapunov exponent for a class of binary
quasiperiodic tight-binding chains, vanishes on the spectrum, which rules
out the presence of localized states. The Fibonacci sequence does not
however belong to this class of chains, and the characterization of the
states in that case remains under discussion. A study on localization
lenghts of tight-binding electrons on a pure Fibonacci chain has been
presented by Capaz \textsl{et al}.,$^{21}$ that found no evidence for
localization of the states.

Real systems have in general some disorder. Random quasicrystals, in the
sense of a random tiling, have been considered$^{22}$ to explain the
properties of quasicrystalline alloys. It is well known that disorder has
pronounced effects on the transport properties of periodic systems,
specially in one-dimension where all the states turn to localized whatever
the amount of disorder. A striking property of quasicrystals is that they
exhibit extremely high resistivities, which decrease with the amount of
defects.$^{23}$ The effects of some types of disorder on the electronic
spectra and wavefunctions of Fibonacci chains have been considered.$^{24-26}$

In this work we study the Lyapunov exponent for electron and phonon
excitations in pure and random Fibonacci quasicrystal chains. We consider
electrons in a tight-binding model, with ''diagonal''-site and
''off-diagonal''-bond Fibonacci ordering, and phonons on a lattice with bond
Fibonacci ordering. The disorder introduced is random tiling imposed on the
substitution or concatenation rules for construction of the Fibonacci
chains. We use a real space renormalization group method, which allows the
calculation of a wavefunction along the chain, for any given energy, and
therefore enables the determination of its Lyapunov exponent. This method
provides a simple and very efficient way of numerically calculating the
Lyapunov exponent as a function of the energy, for large Fibonacci chains.
The method has great similarities, but also important differences, as will
be discussed, with that used by Capaz \textsl{et al.}$^{21}$ The method is
based on decimation, which is here applied either to substitution$^{14}$ or
concatenation,$^{25}$ and implemented in the presence of disorder.

In order to calculate an eigenstate, one needs to specify an energy on the
spectrum. Since the spectrum of a Fibonacci chain is a Cantor set with zero
Lebesgue measure, the probability of numerically specifying an energy on the
spectrum is essentially zero. Hence any chosen energy will almost certainly
correspond to a gap, and the calculated Lyapunov exponents are then
associated to gap states. It is shown that the Lyapunov exponent for the gap
states of the various excitations has a fractal struture, and we study its
scaling properties. From these properties we obtain information on the
Lyapunov exponent for the states on the spectrum of the Fibonacci chain, and
therefore on the localization properties of the excitations. We study the
Lyapunov exponent for both, tight-binding electrons and phonons, remarking
that the Goldstone symmetry present in the later and absent in the former,
may lead to important differences in the scaling properties of the two
systems.

The outline of the paper is as follows. In Sec. II we describe the
tight-binding electron and phonon systems that are studied, and present the
renormalization method used to calculate the Lyapunov exponent. In Sec. III
we present the Lyapunov exponent for the various excitations on a pure
Fibonacci chain, study its scaling properties and discuss localization, and
finally consider the effects of disorder on the Lyapunov exponent. In Sec.
IV we present our conclusions.

\section{Renormalization approach}

The dynamics of tight-binding electron and phonon excitations on a Fibonacci
quasicrystal chain, can be described by the generic equation

\begin{equation}
(\varepsilon _{n}-E)\Psi _{n}=V_{n-1}\Psi _{n-1}+V_{n}\Psi _{n+1}.
\label{(1)}
\end{equation}

\noindent For the electrons, $\Psi _{n}$ denotes the amplitude of the
wavefunction at site $n$, corresponing to energy $E$, $\varepsilon _{n}$ is
a site energy, and $V_{n}$ is the hopping amplitude between site $n$ and 
$n+1 $. For phonons, $\Psi _{n}$ represents the displacement from the
equilibrium position of the atom at site $n$, $E=m\omega ^{2}$, $\omega $
being the phonon frequency and $m$ the atom mass, 
$\varepsilon_{n}=V_{n-1}+V_{n}$, and $V_{n}$ is the spring constant 
connecting sites $n$
and $n+1$. This latter model describes equally well spin waves on an
Heisenberg ferromagnet at zero temperature, replacing the spring constants
by exchange constants, and $m\omega ^{2}$ by the spin wave frequency $\omega 
$. We note the Goldstone symmetry present in the phonon system, which
imposes a correlation between the site $\varepsilon _{n}$ and the coupling 
$V_{n}$ parameters, that is not present in the electron system.

The various Fibonacci quasicrystal models are defined as follows. For
electrons, the ''diagonal'' model is obtained from $\left( 1\right) $ by
setting, $V_{n}=1$ and $\varepsilon _{n}=\varepsilon _{A}$ or $\varepsilon
_{B}$, according to the Fibonacci sequence, and the ''off-diagonal'' model
is obtained from $\left( 1\right) $ by setting, $\varepsilon _{n}=0$ and 
$V_{n}=V_{A}$ or $V_{B}$, according to the Fibonacci sequence. The model for
phonons is obtained from $\left( 1\right) $ with the couplings $V_{n}=V_{A}$
or $V_{B}$, arranged in the Fibonacci sequence.

The disordered Fibonacci chains are built by introducing random tiling in
the substitution rule for construction,

\begin{eqnarray}
B &\rightarrow & A,  \nonumber \\
A &\rightarrow & AB,\hspace{.3cm}\mathrm{  probability}\hspace{.3cm}p,
  \label{(2)} \\
A &\rightarrow &BA,\hspace{.3cm}\mathrm{  probability }\hspace{.3cm}1-p,
  \nonumber
\end{eqnarray}

\noindent in each iteration $i$, starting with $B$, the two possibilities
corresponding respectively to direct and inverse substitution, or they are
built by introducing random tiling in the concatenation rule for
construction,

\begin{eqnarray}
S_{i} &=&S_{i-1}\otimes S_{i-2},\hspace{.3cm}\mathrm{  probability  }
\hspace{.3cm}p
,  \label{(3)} \\
S_{i} &=&S_{i-2}\otimes S_{i-1},\hspace{.3cm}\mathrm{  probability }
\hspace{.3cm}1-p,  \nonumber
\end{eqnarray}

\noindent starting with $S_{0}=B$ and $S_{1}=A$, the two possibilities
corresponding respectively to direct and inverse concatenation. Random
tiling on substitution or concatenation generates, at each iteration, an
identical set of disordered Fibonacci chains, though throug a different
sequence of preceeding chains (e.g. $A\rightarrow AB\rightarrow
ABA\rightarrow ABABA$, by substitution, \textsl{vs},\textsl{\ }
$A\rightarrow BA\rightarrow AAB\rightarrow ABABA$, by concatenation).

The method that we use to calculate the Lyapunov exponent is based on the
fact that the wavefunction $\Psi _{n}$ at the Fibonacci sites 
$n=n(i)=F_{i+1} $, given by $F_{i+1}=F_{i}+F_{i-1}$ with $F_{1}=F_{0}=1$, 
can be easily obtained via a real-space renormalization group transformation,
which consists in eliminating appropriated sites on the chain, so that a
chain similar to the original one is obtained, with a rescaled length and
renormalized parameters. Under successive decimations one carries the system
through larger length scales separating the sites. For the Fibonacci chain
it is possible to deduce an exact renormalization transformation for the
parameters $\varepsilon _{n}$ and $V_{n}$, the rescaling factor, under which
the system is self-similar, being equal to $\tau $. After $i$ iterations,
the renormalization transformation takes, for example, $V_{A}$ to 
$V_{A}^{(i)}$, which represents the renormalized interaction between two
sites that are a distance $\tau ^{i}$ apart, measured in units of the
original lattice spacing. The Fibonacci sites $n(i)$ become the first
neighbours of the end site $n=0$, at each iteration $i$. Now, writing 
$\left( 1\right) $ as a recursion relation for the wavefunction, and fixing
the ''free-end\textit{''} boundary condition $V_{-1}=0$, one gets, 
$\Psi_{1}=\left[ \left( \varepsilon _{0}-E\right) /V_{0}\right] \Psi _{0}$.
 The wavefunction $\Psi _{n}$ at the consecutive Fibonacci sites $n(i)$, can
therefore be obtained in terms of the parameters under successive
renormalization iterations $i$, through

\begin{equation}
\Psi _{n(i)}=\left[ \left( \varepsilon _{0}^{(i)}-E\right)
/V_{0}^{(i)}\right] \Psi _{0},  \label{(4)}
\end{equation}
fixing $\Psi _{0}$ (e.g. $\Psi _{0}=1$). The Lyapunov exponent $\gamma $ is
then calculated from the wavefunction $\left( 4\right) $, given that

\begin{equation}
\left| \Psi _{n}\right| \sim e^{\gamma x_{n}}\mathrm{,\quad }(n\rightarrow
\infty ),\mathrm{ }  \label{(5)}
\end{equation}
and $x_{n}=\tau ^{i}$ for $n=n(i)$. In the work of Capaz \textsl{et al.}
$^{21}$ the localization of the wavefunction $\Psi $ is studied following the
behavior of the coupling $V$ under successive renormalizations, and not
through the evolution of the wavefunction $\left( 4\right) $, which also
involves the parameter $\varepsilon $. Although the behaviour of $\Psi $ is
mainly determined by $V$, the complete expression should be used.
Furthermore, in that work a small imaginary part $\eta $ is added to the
energy $E$, which produces an artificial decay of the coupling $V$, that
alters the actual evolution of the wavefunction, and consequently interferes
in the study of localization and evaluation of the Lyapunov exponent.

Now we present the decimation techniques used to obtain the renormalization
transformation of the parameters $\varepsilon _{0}$ and $V_{0}$, for chains
constructed by substitution or concatenation.

\bigskip

\begin{center}
A. Substitution chains

\medskip
\end{center}

The renormalization transformation of the parameters is obtained by
eliminating sites in such a way as to reverse the substitution procedure in
$\left( 2\right) $.$^{14}$ In order to build the transformation one needs to
consider an expanded parameter space, for the various excitations, where the
bonds $V_{n}$ assume two different values, $V_{A}$ and $V_{B}$, arranged in
a Fibonacci sequence, and the site energies $\varepsilon _{n}$ may assume
three different values, depending on the local environment of $n$, 
$\varepsilon _{\alpha }$ if $V_{n-1}=V_{n}=V_{A}$, 
$\varepsilon _{\beta }$ if 
$V_{n-1}=V_{A}$ and $V_{n}=V_{B}$, $\varepsilon _{\gamma }$ if 
$V_{n-1}=V_{B} $ and $V_{n}=V_{A}$. A choice of the initial parameters $V_{A}$
, $V_{B}$, $\varepsilon _{\alpha }$, $\varepsilon _{\beta }$, $\varepsilon
_{\gamma }$, casts the problem into the model for electron excitations,
''diagonal'' ($V_{A}=V_{B}$, $\varepsilon _{\alpha }=\varepsilon _{\gamma
}\neq \varepsilon _{\beta }$) or ''off-diagonal'' ($V_{A}\neq V_{B}$, 
$\varepsilon _{\alpha }=\varepsilon _{\beta }=\varepsilon _{\gamma }$), or
phonon excitations ($V_{A}\neq V_{B}$, $\varepsilon _{\alpha }=2V_{A}$, 
$\varepsilon _{\beta }=\varepsilon _{\gamma }=V_{A}+V_{B}$). The reversal of
rule $\left( 2\right) $ is achieved through the elimination of $\beta -sites$
, corresponding to direct substitution, or $\gamma $-sites, corresponding to
inverse substitution. The resulting renormalization equations are:

i) direct substitution,

\begin{eqnarray}
\varepsilon _{\alpha }^{(i+1)} &=&\varepsilon _{\gamma }^{(i)}-\left(
V_{A}^{(i)2}+V_{B}^{(i)2}\right) /\left( \varepsilon _{\beta
}^{(i)}-E\right) ,  \nonumber \\
\varepsilon _{\beta }^{(i+1)} &=&\varepsilon _{\gamma
}^{(i)}-V_{B}^{(i)2}/\left( \varepsilon _{\beta }^{(i)}-E\right) ,
\label{(6)} \\
\varepsilon _{\gamma }^{(i+1)} &=&\varepsilon _{\alpha
}^{(i)}-V_{A}^{(i)2}/\left( \varepsilon _{\beta }^{(i)}-E\right) ,
  \nonumber
\\
V_{A}^{(i+1)} &=&V_{A}^{(i)}V_{B}^{(i)}/\left( \varepsilon _{\beta
}^{(i)}-E\right) ,\qquad V_{B}^{(i+1)}=V_{A}^{(i)},  \nonumber
\end{eqnarray}

\noindent and for the end site $n=0$,

\begin{equation}
\varepsilon _{0}^{(i+1)}=\varepsilon _{0}^{(i)}-V_{A}^{(i)2}/\left(
\varepsilon _{\beta }^{(i)}-E\right) ,\quad V_{0}^{(i+1)}=V_{A}^{(i+1)},
\label{(7)}
\end{equation}

ii) inverse substitution, 
\begin{eqnarray}
\varepsilon _{\alpha }^{(i+1)} &=&\varepsilon _{\beta }^{(i)}-\left(
V_{A}^{(i)2}+V_{B}^{(i)2}\right) /\left( \varepsilon _{\gamma
}^{(i)}-E\right) ,  \nonumber \\
\varepsilon _{\beta }^{(i+1)} &=&\varepsilon _{\alpha
}^{(i)}-V_{A}^{(i)2}/\left( \varepsilon _{\gamma }^{(i)}-E\right) ,
\label{(8)} \\
\varepsilon _{\gamma }^{(i+1)} &=&\varepsilon _{\beta
}^{(i)}-V_{B}^{(i)2}/\left( \varepsilon _{\gamma }^{(i)}-E\right) , 
\nonumber \\
V_{A}^{(i+1)} &=&V_{A}^{(i)}V_{B}^{(i)}/\left( \varepsilon _{\gamma
}^{(i)}-E\right) ,\qquad V_{B}^{(i+1)}=V_{A}^{(i)},  \nonumber
\end{eqnarray}

\noindent and for the end site $n=0$,

\begin{eqnarray}
\varepsilon _{0}^{(i+1)} &=&\varepsilon _{0}^{(i)}-V_{B}^{(i)2}/\left(
\varepsilon _{\gamma }^{(i)}-E\right) ,\quad
V_{0}^{(i+1)}=V_{A}^{(i+1)},\quad \mathrm{if }\hspace{.3cm}
V_{0}^{(i)}=V_{B}^{(i)}, 
\nonumber \\
\varepsilon _{0}^{(i+1)} &=&\varepsilon _{0}^{(i)},\quad
V_{0}^{(i+1)}=V_{B}^{(i+1)},\quad \mathrm{if }\hspace{.3cm}
V_{0}^{(i)}=V_{A}^{(i)}.\mathrm{ }
\label{((9))}
\end{eqnarray}

\bigskip

\begin{center}
B. Concatenation chains

\bigskip
\end{center}

The renormalization transformation of the parameters is obtained by
eliminating the central site, after having performed concatenation, so as to
reverse the concatenation procedure $(3)$.$^{25}$ This leads to the
following renormalization equations, which are different for bond Fibonacci
ordering, i.e. ''off-diagonal'' electrons and phonons, or site Fibonacci
ordering, i.e. ''diagonal'' electrons.

For the bond problem:

i) direct concatenation,

\begin{eqnarray}
\varepsilon _{0}^{(i+1)} &=&\varepsilon _{0}^{(i)}-V_{0}^{(i)2}/\left(
\varepsilon _{cd}^{(i+1)}-E\right) ,  \nonumber \\
\varepsilon _{F_{i+1}}^{(i+1)} &=&\varepsilon
_{F_{i-1}}^{(i-1)}-V_{0}^{(i-1)2}/\left( \varepsilon _{cd}^{(i+1)}-E\right) ,
\label{(10)} \\
V_{0}^{(i+1)} &=&V_{0}^{(i)}V_{0}^{(i-1)}/\left( \varepsilon
_{cd}^{(i+1)}-E\right) ,  \nonumber
\end{eqnarray}

\noindent with, $\varepsilon _{cd}^{(i+1)}=\varepsilon
_{F_{i}}^{(i)}+\varepsilon _{0}^{(i-1)}$,

ii) inverse concatenation,

\begin{eqnarray}
\varepsilon _{0}^{(i+1)} &=&\varepsilon _{0}^{(i-1)}-V_{0}^{(i-1)2}/\left(
\varepsilon _{ci}^{(i+1)}-E\right) ,  \nonumber \\
\varepsilon _{F_{i+1}}^{(i+1)} &=&\varepsilon
_{F_{i}}^{(i)}-V_{0}^{(i)2}/\left( \varepsilon _{ci}^{(i+1)}-E\right) ,
\label{(11)} \\
V_{0}^{(i+1)} &=&V_{0}^{(i)}V_{0}^{(i-1)}/\left( \varepsilon
_{ci}^{(i+1)}-E\right) ,  \nonumber
\end{eqnarray}
with, $\varepsilon _{ci}^{(i+1)}=\varepsilon _{F_{i-1}}^{(i-1)}+\varepsilon
_{0}^{(i)}$, and the initial values, $V_{0}^{(0)}=V_{B}$, $V_{0}^{(1)}=V_{A}$
, $\varepsilon _{0}^{(0)}=\varepsilon _{1}^{(0)}=\varepsilon
_{0}^{(1)}=\varepsilon _{1}^{(1)}$, for ''off-diagonal'' electrons, and 
$\varepsilon _{0}^{(0)}=\varepsilon _{1}^{(0)}=V_{0}^{(0)}=V_{B}$, 
$\varepsilon _{0}^{(1)}=\varepsilon _{1}^{(1)}=V^{(1)}=V_{A}$, for phonons.

For the site problem:

i) direct concatenation,

\begin{eqnarray}
\varepsilon _{0}^{(i+1)} &=&\varepsilon _{0}^{(i)}-V_{0}^{(i)2}/\left[
\left( \varepsilon _{F_{i}}^{(i)}-E\right) -T^{2}/\left( \varepsilon
_{0}^{(i-1)}-E\right) \right] ,  \nonumber \\
\varepsilon _{F_{i+1}}^{(i+1)} &=&\varepsilon
_{F_{i-1}}^{(i-1)}-V_{0}^{(i-1)2}/\left[ \left( \varepsilon
_{0}^{(i-1)}-E\right) -T^{2}/\left( \varepsilon _{F_{i}}^{(i)}-E\right)
\right] ,  \label{(12)} \\
V^{(i+1)} &=&TV_{0}^{(i)}V_{0}^{(i-1)}/\left[ \left( \varepsilon
_{0}^{(i-1)}-E\right) \left( \varepsilon _{F_{i}}^{(i)}-E\right)
-T^{2}\right] ,  \nonumber
\end{eqnarray}

ii) direct concatenation,

\begin{eqnarray}
\varepsilon _{0}^{(i+1)} &=&\varepsilon _{0}^{(i-1)}-V_{0}^{(i-1)2}/\left[
\left( \varepsilon _{F_{i-1}}^{(i-1)}-E\right) -T^{2}/\left( \varepsilon
_{0}^{(i)}-E\right) \right] ,  \nonumber \\
\varepsilon _{F_{i+1}}^{(i+1)} &=&\varepsilon
_{F_{i}}^{(i)}-V_{0}^{(i)2}/\left[ \left( \varepsilon _{0}^{(i)}-E\right)
-T^{2}/\left( \varepsilon _{F_{i-1}}^{(i-1)}-E\right) \right] ,  \label{(13)}
\\
V^{(i+1)} &=&TV_{0}^{(i)}V_{0}^{(i-1)}/\left[ \left( \varepsilon
_{0}^{(i)}-E\right) \left( \varepsilon _{F_{i-1}}^{(i-1)}-E\right)
-T^{2}\right] ,  \nonumber
\end{eqnarray}

\noindent with the initial values, $V_{0}^{(2)}=T$, $\varepsilon
_{0}^{(3)}=\varepsilon _{2}^{(3)}=\varepsilon _{A}-T^{2}/\left( \varepsilon
_{B}-E\right) $, $V_{0}^{(3)}=T^{2}/\left( \varepsilon _{B}-E\right) $, and
in i) $\varepsilon _{0}^{(2)}=\varepsilon _{A}$, $\varepsilon
_{1}^{(2)}=\varepsilon _{B}$, while in ii) $\varepsilon
_{0}^{(2)}=\varepsilon _{B}$, $\varepsilon _{1}^{(2)}=\varepsilon _{A}$.

\medskip

Considering the general case of a random Fibonacci chain, for a given
probability of disorder $p$, we start with a specific disordered
configuration, generated by $\left( 2\right) $ or $(3)$, respectively for
substitution or concatenation chains, and then iterate $\left( 6\right) $ 
$-(9)$, $(10)-(11)$ or $(12)-(13)$, depending on the system studied,
according to that configuration, in order to obtain the successive values
for $V_{0}^{(i)}$ and $\varepsilon _{0}^{(i)}$. This allows us to calculate
the wavefunction $\Psi _{n}$, provided by $\left( 4\right) $, at the
successive Fibonacci sites, for a given energy $E$. For each probability 
$p$, we average the obtained wavefunction for $E$ over many different disorder
configurations. It is important to remark that when dealing with random
chains, one should first calculate the wavefunction for a specific
disordered configuration and then average over configurations, instead of
averaging the parameters over disorder at each step of the renormalization
and then calculate the wavefunction with the averaged parameters. This
latter procedure$^{25}$ will wash out important correlations in the system,
and leads to different results depending on how the average is performed.
The first procedure describes the physics more accurately.

\section{Lyapunov Exponent for Fibonacci Chains}

We now present the results concerning the Lyapunov exponent, calculated as a
function of the energy, for the tight-binding electron, ''diagonal'' and
''off-diagonal'', and phonon excitations on the pure and random Fibonacci
chains. We consider first the case of pure chains, for which we study the
scaling properties of the Lyapunov exponent and their implications for the
localization of states on the spectrum, and analyse afterwards the effects
of disorder, of the kind of random tiling, on the Lyapunov exponent.

As mentioned above, the wavefunctions that we numerically calculate
correspond to gap states. Figure 1 shows the typical behavior of a
wavefunction $\Psi _{n}$, at any chosen energy $E$, either for the electron
or the phonon excitations on a pure Fibonacci chain. One observes that the
wavefunction first oscillates over a certain length, and then grows
exponentially. This behavior has mixed characteristics of an extended
(oscillating) state and a localized (exponential) state. The length over
which a wavefunction oscillates is a ''memory'' length,$^{10}$ in the sense
that beyond this length it loses memory of its initial phase. The
exponential growth of the wavefunction is characterized by the Lyapunov
exponent, which measures the inverse of a ''localization length''. We find
that the ''memory'' length $\xi $ and the Lyapunov exponent $\gamma $ are
simply related, $\xi \approx 1/\gamma $. In figure 2 we present the Lyapunov
exponent for the electron, ''diagonal'' and ''off-diagonal'', and phonon
excitations on the pure Fiboncci chain, calculated as a function of the
energy. The exponent exhibits a rather nontrivial dependence on the energy,
which has a clear correspondence with the associated density of states
obtained by Ashraff and Stinchcombe$^{14,15}$ for the various cases, the
finite values of the Lyapunov exponent corresponding to gap states, with the
further a state is inside a gap the larger is its Lyapunov exponent. The
Lyapunov exponent exhibits a fractal structure, i.e. under dilation the same
structure is revealed in a smaller scale, as can be seen by comparing the
Lyapunov plots in figure 2 with those in figure 3. This structure is
observed even in the very low energy range of the magnetic excitations,
where $\gamma $ takes particularly small values, most probably due to the
Goldstone symmetry.

The scaling behavior of the Lyapunov exponent is studied through the
variation of the maximum exponent in a gap, $\gamma _{\max }$, versus the
gap width, $\Delta E_{g}$.$^{21}$ We find that

\begin{equation}
\gamma _{\max }\sim (\Delta E_{g})^{\delta },  \label{((14))}
\end{equation}

\noindent where the scaling index $\delta $, is independent of the energy
for the electron excitations, ''diagonal'' and ''off-diagonal'', as shown in
figure 4, but depends on the energy for the phonon excitations, as figure 5
reveals, and it is shown in figure 6. We also find that the scaling index
for the electron excitations depends on the quasicrystal site ($\varepsilon
_{A}$, $\varepsilon _{B}$) or bond ($V_{A}$,$V_{B}$) parameters, decreasing
as the difference between the parameters increases, while the scaling index
for the phonon excitations, varying with energy, also depends on the
quasicrystal parameters ($V_{A}$,$V_{B}$). Our results for the electron
excitations are in agreement with those obtained by Capaz et al.,$^{21}$
though their scaling indices differ from ours, probably due to the fact that
 they have
calculated the Lyapunov exponent from the behavior of the coupling $V$ alone
and not from the evolution of the wavefunction $\Psi$, in $(4)$, and moreover
have introduced an imaginary part in the energy which influences the 
Lyapunov exponent, as discussed earlier. 

From the scaling expression $\left( 14\right) $, one obtains, for the
various excitations, that $\gamma _{\max }\rightarrow 0$ when 
$\Delta E_{g}\rightarrow 0$, implying that the Lyapunov exponent for 
wavefunctions on the spectrum, vanishes. We therefore have that the 
electron, ''diagonal'' or ''nondiagonal'', and phonon excitations on
 a Fibonacci chain are nonlocalized.

Let us now study the effects of disorder on the Lyapunov exponent. Disorder
has drastic effects on the wavefunctions of one-dimensional periodic
systems, localizing all the states. Figure 7 illustrates this fact, showing
the Lyapunov exponent for phonon excitations on a random periodic chain,
with couplings $V_{A}$ and $V_{B}$, as a function of the probability $p$ of
disorder, for various energies. One sees that the Lyapunov exponent
increases with disorder, being also an increasing function of the energy.

For the random Fibonacci chains we considered disorder of the kind of random
tiling, introduced in the substitution or concatenation rule for
construction of the chains. The resulting disordered chains differ from the
pure Fibonacci chain in having a varying number of phason flips, located at
certain points on that chain. By a phason flip it is meant a local
rearrangement of tiles on the quasiperiodic structure, corresponding to a
switch of the site, $\varepsilon _{A}$ and $\varepsilon _{B}$, or the bond, 
$V_{A}$ and $V_{B}$, parameters.$^{26}$ Using the cyclic property of the
trace one can see that all those random tiling chains have the same
spectrum, for the electron and the phonon excitations, as the pure Fibonacci
chain. In the work of L\'{o}pez \textsl{et al}.,$^{25}$ on the effects of
that kind of random tiling on the electronic excitations of a Fibonacci
chain, it has however been found that the disorder affects the spectrum of
the excitations. We think that this result is a consequence of the averaging
of the parameters over disorder taken at each step of the renormalization in
that work, which as already mentioned, loses important correlations in the
system and introduces effects that depend on the averaging procedure used,
corresponding in fact to different systems. On the other hand Naumis and
Arag\'{o}n,$^{26}$ considering electronic excitations, have also noted that
phason flips located at certain points on the Fibonacci chain do not alter
the spectrum of the excitations.

We calculated the Lyapunov exponent for the electron and phonon excitations
on Fibonacci chains with random tiling, as a function of the probability of
disorder, for different values of energy. The results obtained are
illustrated in figure 8. We find that the disorder considered does not
affect the Lyapunov exponent either for the electron, ''diagonal'' or
''off-diagonal'', or the phonon excitations. The same result is obtained for
disordered Fibonacci chains with random tiling either in the substitution or
the concatenation rule for construction of chains. The irrelevance of
disorder found for the Lyapunov exponent of excitations on a Fibonacci chain
is surprising, knowing the drastic effects that disorder has on the
excitations on periodic chains. However, it should be noted that random
tiling introduces a kind of bounded disorder, which has also correlations,
and therefore might not be sufficient to produce localization of states.
Furthermore, in contrast to the general case, it has been reported that
there exist particular random potentials in one dimension that allow for
extended states, those being described by an iterative procedure of
construction.$^{27,28}$ Liu and Riklund$^{24}$ have found that other types
of disorder, different from the one considered by us, produce localization
of electronic excitations on a Fibonacci chain.

\section{Conclusions}

We have studied the Lyapunov exponent for tight-binding electron,
''diagonal'' and ''off-diagonal'', and phonon excitations in pure and random
Fibonacci quasicrystal chains, using a real space renormalization group
method. This method allows the calculation of a wavefunction along the
chain, and the determination of the associated Lyapunov exponent, as a
function of the energy, in a very efficient way for very long chains. We
have found that the Lyapunov exponent for the pure Fibonacci chain has a
self-similar structure, being characterized by a scaling index that is
independent of the energy for the electronic excitations, but depends on the
energy for the phonon excitations. The scaling properties of the Lyapunov
exponent, imply that it vanishes on the spectrum for the various
excitations. We therefore have that the electronic and phonon excitations
are nonlocalized on the Fibonacci chain . Considering random Fibonacci
chains, we calculated the Lyapunov exponent as a function of the probability
of disorder, and found that the disorder introduced, of the kind of random
tiling, does not affect the Lyapunov exponent, which takes the same value as
for the pure Fibonacci chain whatever the degree of disorder, either for the
electron or for the phonon excitations. The random tiling considered
generates in fact chains that are locally isomorphic to the pure Fibonacci
chain, and therefore our results imply that locally isomorphic chains,
besides having the same energy spectrum,$^{29}$ also have the same Lyapunov
exponent, and hence their eigenstates have the same nature as the ones of
the pure Fibonacci chain. We are now investigating the effects of random
tiling on the Lyapunov exponent, of electron and phonon excitations, on
other aperiodic chains, such as the Thue-Morse, the period-doubling, and
binary non-Pisot sequences. Other types of disorder are also being
considered on the Fibonacci chain, as well as on the other aperiodic chains
mentioned, in order to understand the relevance/irrelevance of disorder on
the Lyapunov exponent, and consequently on the localization properties of
those systems. The results of this work will be reported elsewhere.

\bigskip

\noindent {\Large Acknowledgements}

We would like to thank R. B. Stinchcombe and J. M. Luck for very helpful
conversations.

\newpage

\noindent {\Large References}

\bigskip

\noindent $^{\dagger }$iveta@alf1.cii.fc.ul.pt

\noindent $^{1}$D. Shechtman, I. Blech, D. Gratias and J. W. Cahn, Phys.
Rev. Lett. \textbf{53}, 1951 (1984).

\noindent $^{2}$R. Merlin, K. Bajena, R. Clarke, F. Y. Juang, and P. K.
Bhattacharya, Phys. Rev. Lett. \textbf{5}, 1768 (1985).

\noindent $^{3}$P. J. Steinhardt and S. Ostlund, \textit{The Physics of
Quasicrystals} (World Scientific, Singapore, 1987).

\noindent $^{4}$D. P. DiVicenzo and P. Steinhardt, \textit{Quasicrystals:
The State of the Art} (World Scientific, Singapore, 1991).

\noindent $^{5}$C. Janot, \textit{Quasicrystals - a primer}, 2nd. ed.
(Clarendon Press, Oxford, 1994).

\noindent $^{6}$M. Kohmoto, L. P. Kadanoff and C. Tang, Phys. Rev. Lett. 
\textbf{50}, 1870 (1983).

\noindent $^{7}$S. Ostlund and R. Pandit, Phys. Rev. B \textbf{29}, 1394
(1984).

\noindent $^{8}$M. Kohmoto and Y. Oono, Phys. Lett. \textbf{102A}, 145
(1984).

\noindent $^{9}$M. Kohmoto and J. Banavar, Phys. Rev. B \textbf{34}, 563
(1986).

\noindent $^{10}$J. M. Luck and D. Petritis, J. Stat. Phys. 42, 289 (1986).

\noindent $^{11}$M. Kohmoto, B. Sutherland and C. Tang, Phys. Rev. B 
\textbf{35}, 1020 (1987).

\noindent $^{12}$B. Sutherland, Phys. Rev. B \textbf{35}, 9529 (1987).

\noindent $^{13}$Q. Niu and F. Nori, Phys. Rev. Lett. \textbf{57}, 2057
(1986).

\noindent $^{14}$J. A. Ashraff and R. B. Stinchcombe, Phys. Rev. B 
\textbf{37}, 5723 (1988).

\noindent $^{15}$J. A. Ashraff, D.Phil. Thesis, Oxford 1989.

\noindent $^{16}$A. S\"{u}to, J. Stat. Phys. \textbf{56}, 525 (1989).

\noindent $^{17}$K. Ishii, Supp. Progr. Theor. Phys. \textbf{53}, 77 (1973).

\noindent $^{18}$J. M. Luck, \textit{Syst\`{e}mes D\'{e}sordonn\'{e}s
Unidimensionnels} (Collection Al\'{e}a - Saclay, 1992).

\noindent $^{19}$A. Crisanti, G. Paladin and A. Vulpiani, \textit{Products
of Random Matrices}, Vol. 104 of Springer Series in Solid-State Sciences
(Springer, Berlin, 1993).

\noindent $^{20}$F. Delyon and D. Petritis, Commun. Math. Phys. \textbf{103}
, 441 (1986).

\noindent $^{21}$R. B. Capaz, B. Koiller and S. L. A. Queiroz, Phys. Rev. B
42, 6402 (1990).

\noindent $^{22}$C. L. Henley, in Ref. 4, p. 429.

\noindent $^{23}$K. Kimura and S. Takeuchi, in Ref. 4, p.313.

\noindent $^{24}$Y. Liu and R. Riklund, Phys. Rev. B \textbf{35}, 6034
(1987).

\noindent $^{25}$J. C. L\'{o}pez, G. Naumis and J. L. Arag\'{o}n, Phys. Rev.
B \textbf{48}, 12459 (1993).

\noindent $^{26}$G. G. Naumis and J. L. Arag\'{o}n, Phys. Lett. A 
\textbf{244}, 133 (1998).

\noindent $^{27}$J. S. Denbigh and N. Rivier, J. Phys. C \textbf{12}, L107
(1979).

\noindent $^{28}$A. Crisanti, C. Falesia, A. Pasquarella, and A. Vulpiani,
J. Phys.: Condens. Matter \textbf{1}, 9509 (1989).

\noindent $^{29}$F. Wijnands, J. Phys. A \textbf{22}, 3267 (1989).

\newpage

\noindent {\Large Figure captions}

\medskip

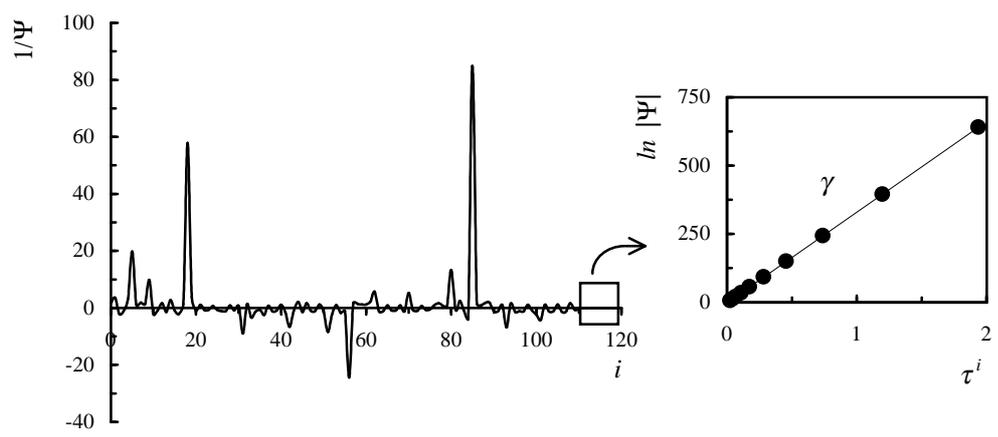
\noindent FIG. 1. Inverse of the phonon wavefunction $\Psi _{n}$ at the
Fibonacci sites $n=F_{i+1}$, for energy $E=0.4$, on a pure Fibonacci chain,
and the associated Lyapunov exponent $\gamma $.

\medskip

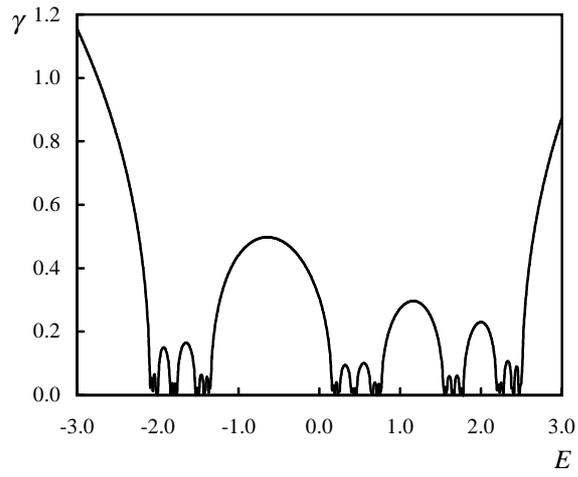
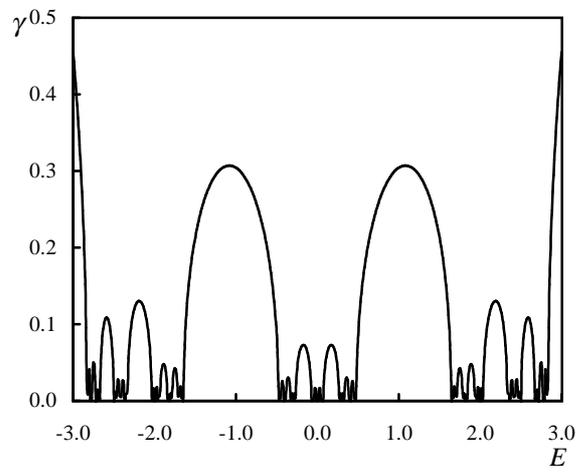
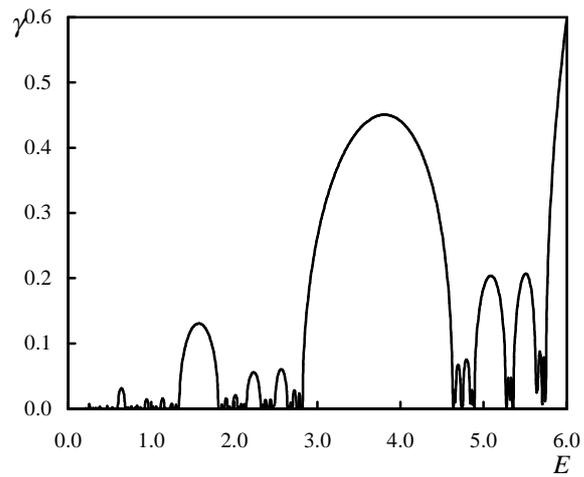
\noindent FIG. 2. Lyapunov exponent $\gamma $ for: a) electronic
''diagonal'' ($\varepsilon _{\alpha }=-\varepsilon _{\beta }=\varepsilon
_{\gamma }=1$), b) electronic ''off-diagonal'' ($V_{A}=1,V_{B}=2$), and c)
phonon ($V_{A}=1,V_{B}=2$) excitations on a pure Fibonacci chain.

\medskip

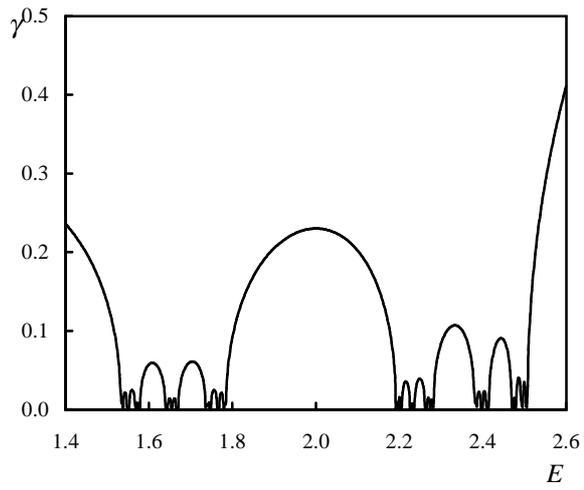
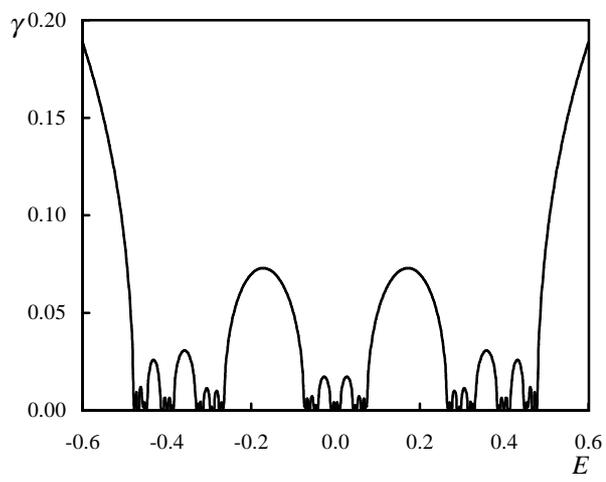
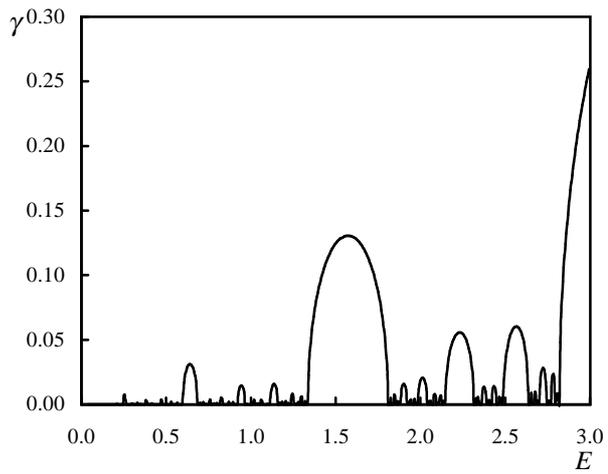
\noindent FIG. 3. Self-similar structure of $\gamma $: a) electronic
''diagonal'' ($\varepsilon _{\alpha }=-\varepsilon _{\beta }=\varepsilon
_{\gamma }=1$), b) electronic ''off-diagonal'' ($V_{A}=1,V_{B}=2$), and c)
phonon ($V_{A}=1,V_{B}=2$) excitations, to compare with FIG. 2.

\medskip

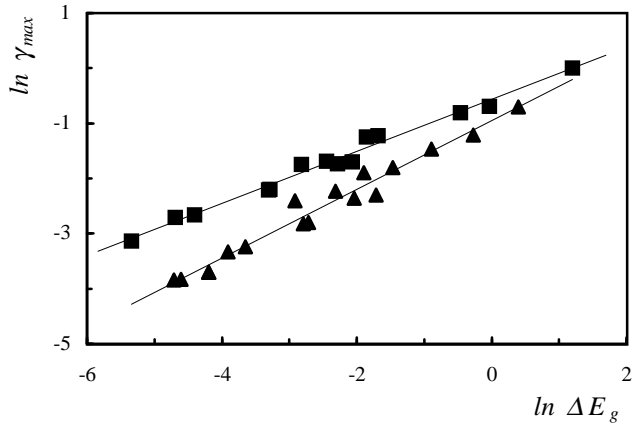
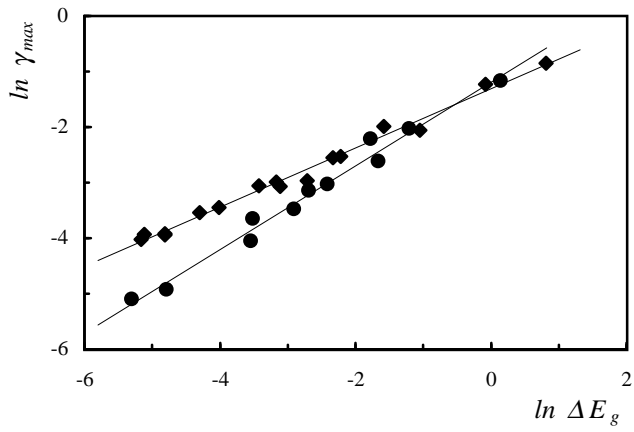
\noindent FIG. 4. Maximum $\gamma $ in a gap \textsl{vs} gap width $\Delta
E_{g}$, for: a) electronic ''diagonal'', ($\blacktriangle $) ($\varepsilon
_{\alpha }=-\varepsilon _{\beta }=\varepsilon _{\gamma }=1$, $\delta =0.62$
), ($\blacksquare $) ($\varepsilon _{\alpha }=-\varepsilon _{\beta
}=\varepsilon _{\gamma }=2$, $\delta =0.47$); b) electronic
''off-diagonal'', ($\bullet $) ($V_{A}=1$, $V_{B}=2$, $\delta =0.75$), 
($\blacklozenge $) ($V_{A}=3$, $V_{B}=1$, $\delta =0.53$), excitations on a
pure Fibonacci chain.

\medskip

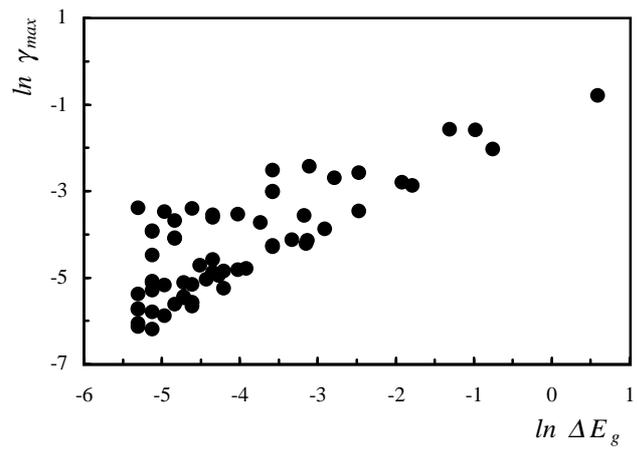
\noindent FIG. 5. Maximum $\gamma $ in a gap \textsl{vs} gap width $\Delta
E_{g}$, for phonon excitations ($V_{A}=1$, $V_{B}=2$), on a pure Fibonacci
chain.

\bigskip

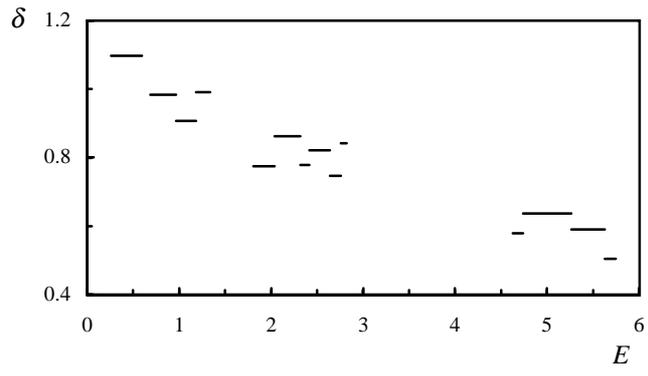
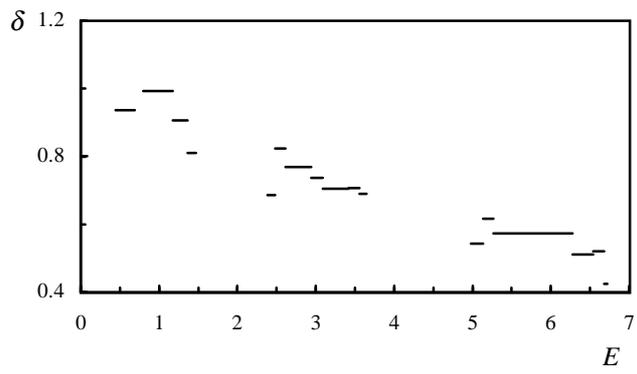
\noindent FIG. 6. Power-law exponent $\delta $, of $\gamma _{\max }$ 
\textsl{vs} $\Delta E_{g}$, for phonon excitations: a) $V_{A}=1$, $V_{B}=2$, b) $V_{A}=2$, $V_{B}=1$, on a pure Fibonacci chain.
\medskip

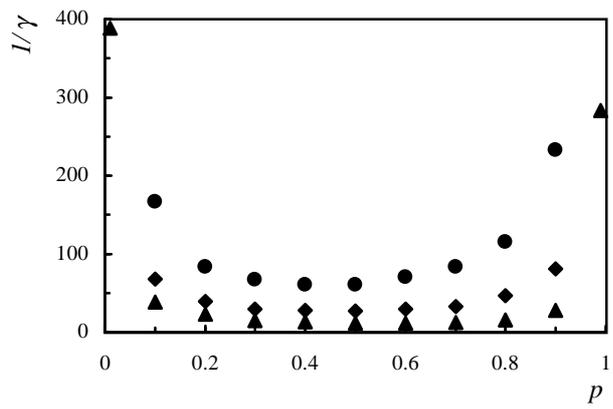
\noindent FIG. 7. Lyapunov exponent $\gamma $ \textsl{vs} probability of
disorder $p$, for phonon excitations, with energies: ($\bullet $) $E=1.2$, 
($\blacklozenge $) $E=2.3$, ($\blacktriangle $) $E=3.4$, on random periodic
chains.

\medskip

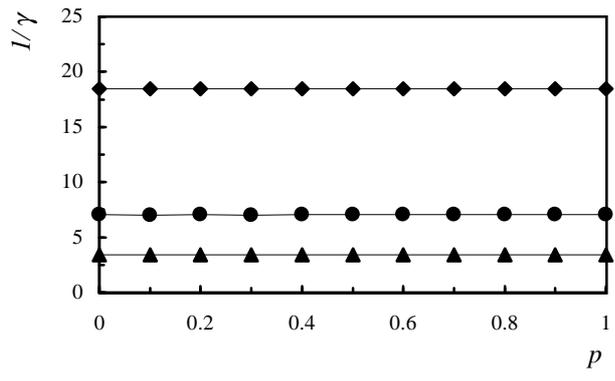
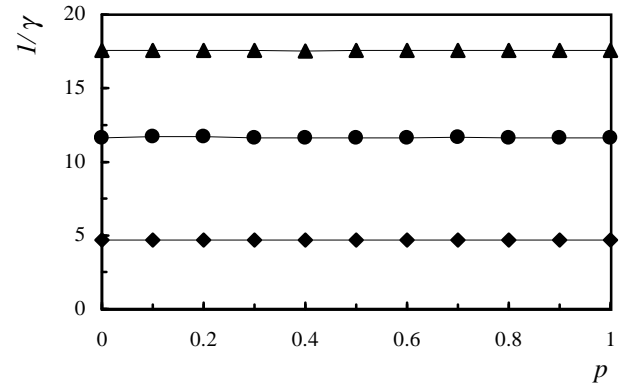
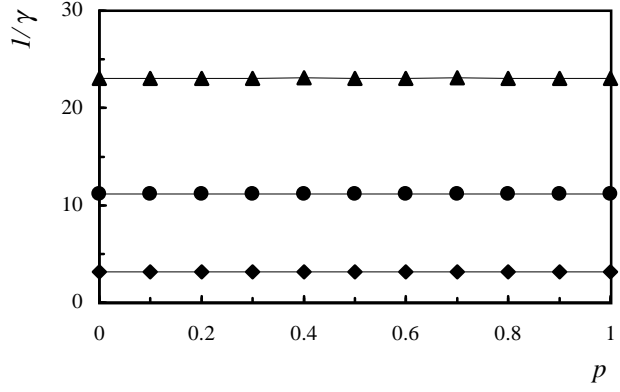
\noindent FIG. 8. Lyapunov exponent $\gamma $ \textsl{vs} probability of
disorder $p$, for various energies $E$, of: a) electronic ''diagonal'', 
($\bullet $) $E=-1.9$, ($\blacklozenge $) $E=0.15$, ($\blacktriangle $) 
$E=1.1$; b) electronic ''off-diagonal'', ($\bullet $) $E=0.5$, 
($\blacklozenge $) $E=1.5$, ($\blacktriangle $) $E=2.05$, c) phonon,
($\bullet $) $E=1.4$, ($\blacklozenge $) $E=3.1$, 
($\blacktriangle $) $E=5.29$, excitations on
random Fibonacci chains.

\end{document}